# "Reminder: please update your details": Phishing Trends

Cynthia Dhinakaran and Jae Kwang lee
Department of Computer Engineering
Hannam University, Daejeon
South Korea

Dhinaharan Nagamalai
Wireilla Net Solutions Inc, India

*Abstract*— **Spam messes up users inbox, consumes resources and spread attacks like DDoS, MiM, Phishing etc., Phishing is a byproduct of email and causes financial loss to users and loss of reputation to financial institutions. In this paper we study the characteristics of phishing and technology used by phishers. In order to counter anti phishing technology, phishers change their mode of operation; therefore continuous evaluation of phishing helps us to combat phishers effectively. We have collected seven hundred thousand spam from a corporate server for a period of 13 months from February 2008 to February 2009. From the collected date, we identified different kinds of phishing scams and mode of their operation. Our observation shows that phishers are dynamic and depend more on social engineering techniques rather than software vulnerabilities. We believe that this study would be useful to develop more efficient anti phishing methodologies.**

## I. INTRODUCTION

Modern world believes that email is great tool for information management and dissemination. E-mail provides easy path for data flow in and out of a controlled local network. At the same time, the market of illegal business through Internet is growing. Phishing is one kind of such business that challenges online transaction systems. Phishing is closely related to spamming. Phishing mail reaches millions of end users by spam[5]. The spammers are using well designed software tools as well as social engineering methodologies to reach end users. The phishing tools are highly sophisticated such as the Phishers can fix the duration of the attack, frequency of attack to the particular target, subject of the phishing mail, contents of the message, hiding the source of the mail etc., Some of the sample automated tools are Email Spoofer, Bulk Mailer, Aneima 2.0, Avalanche 3.5, Euthanasia etc[4]. Phishing is a kind of spam causing reputation, economic and man power loss. Phishers heavily utilize social engineering techniques to lure email users and divulge their valuable data. The data includes name, pass phrase, social security numbers, telephone numbers, address, and email accounts etc. There are many types of Phishing that have been reported in recent years. a) The most common way is to divert the end-user to fraudulent website controlled by a phisher by clicking a hyperlink available in an email. The fraudulent website asks the user to divulge the sensitive information and this leads to loss. B) Another Phishing attack is to ask the user to contact a Phishers telephone number or fax number. C) The third way is thro instant message from unknown contacts that leads to password and identity theft.

Most Phishing activities related to financial institutions, ecommerce websites, and online greeting card services. Phishing is closely related to spamming. We identified that some spammers are doing spamming as well as phishing simultaneously. Anti spam technology ranges from simple content filters to sophisticated software based on various algorithms such us Bayesian and are widely used to stop spam. But spammers always find new ways to reach users inbox. The arms race between anti spam software developers and spammers causes huge concern to the Internet community. Anti spam filters are highly useful to defend the phishing scams. Apart from this there are plenty of tools that are available to fight against the phishing scams. For example CallingId,Cloudmark, Earthlink, ebay, Netcraft, Trustwatch, Spoofguard, site advisor[3] are some of the anti phishing tools worth to mention here. But unfortunately most of these tools do not providing adequate protection to the end users. According to the [3] research the best anti phishing software tool identifies 50% of false positives. There are more than 2 dozen free anti phishing tools that are available to the end users. Since Phishers are changing their tactics regularly based on anti phishers take down approach as well as anti phishing software, the anti phishing tools are not able to cope up with the technology of Phishers. Phishing is an attack exploiting human vulnerabilities rather than technical vulnerabilities. According to gardener [18] survey, phishing will grow in near future, since it is a profitable business. In this paper we will discuss how the phishers do their business successfully and what kind of technology they use and how they are make use of social engineering methods. Finally we conclude with effective countermeasures to this problem.

The rest of the paper is organized as follows. Section 2 provides background on phishing and spam and the effects. Section 3 provides data Collection and experimental results. In section 4&5 we describe the methodologies used by phishers. Section 6 provides the countermeasures to the phishing scam. We conclude in section 7.

## II. RELATED WORK

[1] Authors analysis the human factors involved in phishing attacks and suggest few techniques that can be used as remedy. According to their study, mostly phishing emails are not signed by a person but by position like Account manager, PayPal etc. Phishers are least bothered about the design, spelling errors in their web site and copy right information. Even legitimate web page emphasizes more on security issues, creates panic among the user community. Their study reveals that some of the third party security endorsements are not well received by the users. Also the email stimuli is more phishy than web stimuli and the phishers use automated phone messages to increase the vulnerability of the users. [2]The authors have examined the characteristics of phishing URLs, domain and machines used to host phishing domains. Phishers are more active than ordinary users and they activate their sites immediately after registering their domain and bring it down whenever they want. Phishing URLs normally contain the target's name but the length of the target name is different from the original one. Phishing domain names are significantly different from regular English character frequency. Their study reveals that mostly the phishing hosted servers are not available in standard ports. Phishers usually do this to avoid the identification of life span of the web site. [3] Presented a multi layer approach to defend DDoS attacks caused by spam. Their study also reveals the effectiveness of SURBL, DNSBLs, content filters. They have presented characteristics of virus, worms and Trojans accompanied by spam as an attachment. The multilayer approach is defending DDoS attacks effectively. [4] The authors observed that spammers use automated tools to send spam with attachment. The main features of this software are hiding sender's identity, randomly selecting text messages, identifying open relay machines, mass mailing capability and defining spamming duration. They tabled that, heavy users email accounts are more vulnerable than relatively new email accounts. [5] The authors tested ten well known anti phishing software tools to evaluate the performance of anti phishing software tools. Their study reveals that none of the tools are 100% accurate, only one tool works better than rest with more than 42% of false positive. The authors have also identified that many of the phishing sites were taken down with hours. Some of these tools didn't detect phishing. In our study, we identified that phishers send phishing mails from different sources or different IP addresses. The take down approach works good to chase phishers temporarily but not to eliminate them totally. The authors have identified that the performance of anti phishing tools varied depending on the source of phishing URLs. These tools are based on black lists, white lists, heuristics and community ratings. Most of these technologies are similar to anti spam technologies with little modifications done for phishing.

## III. DATA COLLECTION

We have characterized Phishing mail from a collection of over 700 000 spam over a period of 13 months from Feb 2008 to Feb 2009 from a corporate mail server. The mail server provides services to 280 users with 30 group email accounts and 290 individual mail accounts. The speed of Internet connection is 100 Mpbs, with 40 Mbps upload and download speed (Due to security and privacy concerns we unable to disclose the real domain name).In order to segregate spam from legitimate mails, we have conducted standard spam detection tests on our server. The spam mails detected by these techniques were directed to the spam trap that is set up on mail server. The spam mails were usually related to finance, pharmacy, business promotion, adultery services and viruses. Considerable portion of these spam mails were used for phishing, DDoS attacks, MiM attacks etc.

From this collection, we separated phishing mails by a small program written in python. In this paper we study the characteristics of Phishing and the technology used by phishers. In order to counter anti phishing technology, phishers change their mode of operation frequently, this leads to continues evaluation of the characteristics of Phishing and phishers technology. The results of these evaluations help us to enhance the existing anti phishing technology and combat phishers effectively.

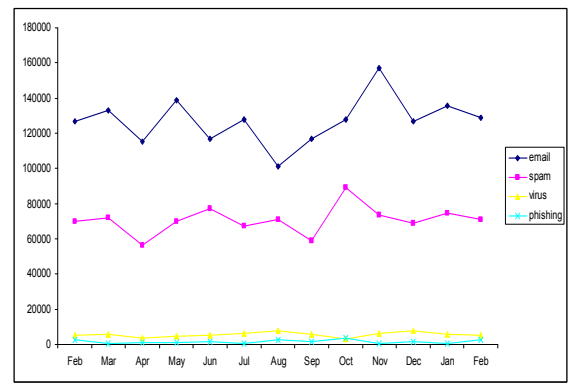

Figure 1. Mail traffic

Figure 1 shows the incoming mail traffic of our mail server for 13 months from Feb 2008 to Feb 2009. X axis is month and y axis is the number of mail, spam, virus, phishing mails received by end users. As seen from the graph, the spam traffic is not related to virus and phishing quantitatively. But our analysis shows common factors in spam and phishing like senders, and the technology used. Roughly the number of phishing mails ranges from 369 to 3459 with an average rate of 1516 per month. This statistics doesn't include false negatives. We believe that this analysis could be useful to develop more efficient anti phishing techniques.

## IV. PHISHING METHODOLOGIES

Since phishers use combination of spam and other techniques to reach end users, we have generated a table of our findings in this section. Our spam collection contains major phishing scams as shown in the table 1. Some other virus, worms, trojon are also used to phish end users. There are two types of phishing methodologies have been detected through our data collection. One method is by using simple link in the spam mail where phishers lure end users to divulge their sensitive information. In the second phishers approach the end users by using trojon, malware & virus. The following are the malware, trojon & virus which are used for phishing.

TABLE I. IDENTIFIED PHISHING SOURCES

| Name of virus, worm, Trojan |
|---|
| Spam.Phish.url |
| Spam.Hoax.HOAX_PHISH_FORGED_PAYPAL |
| Clm.HTML.Phishing.Pay-110 |
| Spam.Hoax.HOAX_PHISH_FORGED_EBAY |
| Spam.Hoax.HOAX_PHISH_FORGED_CITIBNK |
| Trojan.spy.html.bankfraud.od |

### A. spam.phish.url

This is a common phishing scam that is used to target all the types of users with all brands. The major brand spoofed by spam.phish.url are citi, paypal, 53, fifth third bank and some other small brands. The number of spam mail received by these scam are far more than others. As shown in the figure 2. number of such mails 100 times more than any other particular phishing scam as shown in the table.

We have identified two types of phishing methods for spam.phish.url. In the first method the phishers directly send spam to end-users through their own machines with well designed software. By doing so phishers target big brands. In the second method spammers target unpopular brands and its users through botnet. Mostly the target contains login and password page. The users are asked to enter the login and password which are controlled by phishers. Their subject line is not properly designed like the previous one.

### B. Spam.hoax.Hoax_phish_forged_paypall

This phishing spam targets citi bank account holders to divulge their citibank ATM/Debit, card number and PIN that customer use on ATM. The link given in the mail takes the user to a non-secure site controlled by the phishers. Mostly these sites are owned by fraudster with fake names. When we checked these websites, they seemed to have disappeared due to take down approach. Mostly the phishing related to citi bank is short lived and the number of scam our domain received in this category is very less as shown in the figure.

### C. Spam.hoax.Hoax_phish_forged_paypall

This is another kind of phishing scam that targets paypall users. These spam ask user to divulge their paypal identity by using social engineering techniques tells users to update their identity as earlier as possible before doing any activity based on email money transaction. The activities include online purchase, money transfer through email, and payment to the commercial websites. The phishing mail asks users to update their details before using the paypal account. The user will be taken to non secure website controlled by fraudsters.

### D. Clm.HTML.Phishing.Pay-110

This type of phishing scam doesn't target any particular type of institutions or a brand. Instead it targets naive users to disclose credit card numbers, bank account information and various other personal details. These mails pretend to be from legitimate authority and demand the confirmation of personal data. The working principle is similar to other phishing scams. Some of these mails might contain attachments but opening these attachments are not harmful. The mail contents are modified based on the target and users. Clm.HTML.Phishing.Pay-110 is a Trojan horse but not spread automatically like other DDoS attack causing spam. It will also not install any software or modify to the destination systems registry.

### E. Spam.Hoax.HOAX_PHISH_FORGED_EBAY

In this kind of phishing scam targets ebay users. The number scams received by these users are more than other scams as shown in the table. The take down approach of this brand does not completely eliminating the scam. As seen n the table, there are fluctuations in the number of mails received by the users. This follows the same methodology used by other brand spoofers like paypal, citi etc. The figure shows the traffic of all these virus,worms & trojons as below.

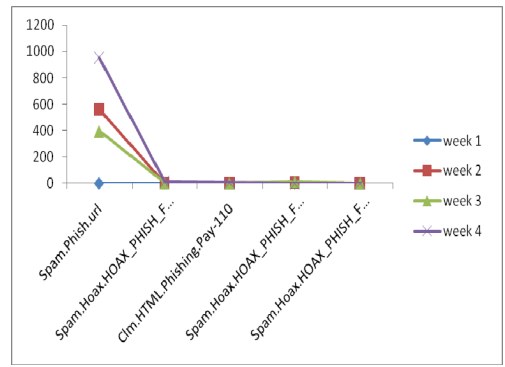

Figure 2. virus,worms & trajons traffic

This is a kind of Trojan virus used for phishing scam. This trojan targets all kinds of users and brands similar to

spam.phish.url. All information such as passwords, credit card details, account details can be stolen using this spam. various version of this scam is also received by end users.

## V. PHISHING TECHNOLOGY & SOCIAL ENGINEERING

Phishers targets end users with extra ordinary precautions based on social engineering techniques. Instead of targeting the entire domain by using brute force method, phishers send mails to small group of email accounts (end users). On the first day, my domain received only 2 unique phishing mails for a particular brand. After that the number of emails that target a particular brand and end users have steadily increased as shown in figure 3. After 4~5 days once again users get reminder mail for their first phishing mail.

Phishers always send reminder mails to make the end users believe that the mail is legitimate. If the phishers use brute force attack for the domain, the scam will reach the entire organization and the end users will be aware of the attack. To avoid this, phishers target small number of users first and day by day they are increasing the number of targets on the same domain

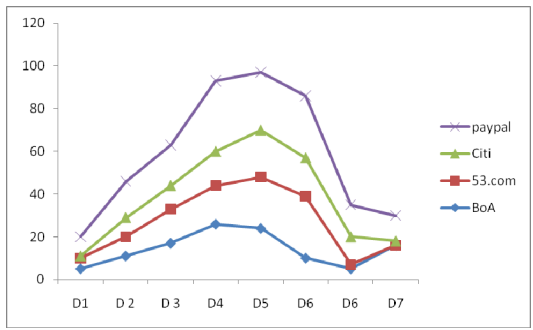

Figure 3. Phishing mails for particular target

The phisher follows the end users regularly. After the first mail, they change their subjects and contact the end user again. The senders email account is used only once. A different email account is used to send reminder or confirmation mail. Usually the confirmation mail address contains the word refid (reference id) just to make the users believe it is a legitimate reminder.

### A. Methodology behind the attack- Charactering the sender ("from") address

The phisher pay lot of attention to social engineering techniques to lure end users. The design of the mail such as sender id, subject line, and contexts are carefully designed to lure the victims. We will be discussing further details. These scam messages are sent by using spam software. These spam program or tools have the facility to generate email accounts as instructed. It can generate fake email accounts, subjects, duration of the spamming, domain name creation and modify the body of the message sent to the users [7]. This spam software can also generate the sender email accounts with a specific format. The senders account plays an important role to lure the victims to divulge the sensitive information. The sender's id designed such as real email accounts from the spoofed brands. The senders email account is designed exactly similar to the format of the legitimate user accounts. The length of the sender account is always more than 21 characters and up to 28 characters. It has three parts before the "at" symbol. The format of the senders mail account is as follows.

(Word1) (numericvalue) (word1 )@ forged domain.com

Word1 contains sensitive words such as customerservice, support, operator, service-number, operator_id, Clientservice,ref, referencenumber, customers, accsupport. From these words we can see that these are people who hold responsible jobs and have access to customer's information. The length of the second numeric part ranges from 5 to 12. In some cases word 1 and numeric word are connected by – or _ symbols as shown in the figure. The length of the third portion is only 2 characters .Mostly they are characters such are ib,ver,ct etc., as shown in the figure 4.

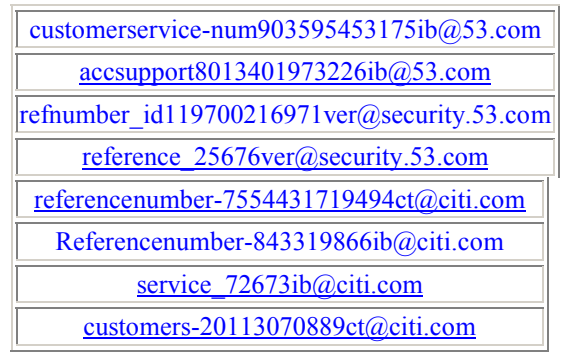

Figure 4. Sender id syntax

### B. Analyzing the "subject" of the phishing mail

Another important way to make user open spam is by attractive subject line. The subject line is carefully designed to lure victim. Examples of attractive subject are "to account confirmation", "message from the bank", "security warning", "update details" etc., By looking at the subject line, the end users think these are legitimate important mails that need to be read. Some of the subject contains date and time stamping to make the end users believe the messages are legitimate. The format of time is day, date month year, time ( hour:minute:seconds-time zone)- The subject line contain only three time zones -0800,-0600,-0500. All of these time zones are from US or Canada. Since all these brands are North American brands, its easier to make the users believe only North American time zones are used. There is no other time zone stamping in phishing mails. The phishing mails subject line can also confirm bank name, phrases like customer service etc., with date and time to make the end-users believe that the message is legitimate. The end users get phishing mails every day in a week.

Mostly phishing mails sent on during the weekends have date and time in the subject area.

| |
|---|
| Fifth Third Bank - confirm your information! |
| Fifth Third Bank - secure confirmation |
| Bank of America - official information! |
| Citibank - Please Confirm Your Information |
| e-banking account confirmation |
| attention from fifth bank |
| security maintenance. |
| reminder: please update your details |

Figure 5. Subject

| |
|---|
| notification from Citibank. -Tue, 13 Jan 2009 08:15:07 -0800 |
| Fifth Third Bank reminder: confirm your account details -Tue, 13 Jan 2009 16:24:33 -0000 |
| Fifth Third Bank - urgent security notification for client. -Tue, 13 Jan 2009 08:31:15 -0800 |
| customer service: your account in Fifth Third Bank. – Tue, 13 Jan 2009 18:13:37 -0600 |
| Alert! [Tue, 13 Jan 2009 11:49:19 -0500] |

Figure 6. Subject with time stamp

### C. Difference between phishing and DDoS attacks

Phishing scams are mass-mailed by group of criminals. Although received by e-mail, they do not spread themselves like DDoS attacks. DDoS attacks also spreads through emails via phishing but DDoS targets the entire domain and causes the service interruption. Phishing targets only financial institutions for financial gain. Phishing mail does not spread from users address book of the infected machines as DDoS attack. DDoS attacks messes up with registry settings results non usability of the system. The infected user machines will download malicious code from the attacker's machine and damage the network of the victim. But the phishing scam doesn't harm the victim's network.

### VI. COUNTERMEASURES

There are more than 100 anti phishing tools freely available to combat phishing threats but none of the anti phishing tools stop phishing effectively. Phishers always come up with new methodlogy to by pass anti phishing countermeasuers. Phishing is heavily dependent on social engineering techniques rather than technological innovations. From our study, we understand that most spammers involved in phishing activities follow the same methodlogies of Spammers. Effective anti spam methods should be implemeted to countermeasures spam threats [4]. Since there are no bullet proof mechanism to combat spam, the combination of methods such as SURBL, DNSBL, rDNS etc and content filter techniques can help stop spam to reach end users inbox. Defending spam will help to reduce the risk of phishing attacks.User awareness also play an important role to avoid be tricked by phishes. But some times bostering more about the security will cause an negative impact on the user community {as mentioned by jacbson}. Finally we conclude by recommemding multi layer spam,phishing defending approach and user awareness are the best way to deal with phishers.Phishers doesn't bring down the network or cause any interruption. It just steals users information for finanacial gain.

### VII. CONCLUSION

Phishing is a by product of spam and data shows that large victims of phishing are end users, brands and financial isntitutions. We have analyzed millions of spam mails received in our server spam trap. Oue analysis show that that there are two type of phishing attack one is by targetting end users by sending emails to divulge their sensitive informtion to the phishers controlled machines by clicking the link. The second category is through virus,malware and trojon which will infect and take user to fradulent websites and to divulge their idendity. Usually spamming techniques are used tor reach end users inbox. Offlate spammers are involved in spamming and spamming attacks. Phishers use social engineering thechniques rather than software vulnerabilites. Phishers usually target users by sending mails to small group in the particular domain and then slowly increase their reach.After a few days, the phsihers start sending "reminder" mails to their initial mail targets to make them believe that the mail is legitimate. If they use brute force method, the phishers think that the user will think it's a hoax mail, which will prevent the phishers out reach. Spammers usually use a brute forec method only to spam end users to mimic DDoS attacks. In the past there is an observation from the study that the number of spam mails go down during the weekend[9]. But recent study [6] reveals that it is not true. Since spammers change their tactice to bypass anti spam countermeasuers, they don't minimize their activity during weekends. There are several specially designed software tools available to automatically generate spam and phisshing mails. So the weekend doesn't matter for the phishers.